\documentclass[twocolumn,amsmath,amssymb,prl]{revtex4}
\usepackage{graphicx}% Include figure files
\usepackage{dcolumn}% Align table columns on decimal point
\usepackage{mathrsfs}
\usepackage{bm,amsmath}
\usepackage{times}
\begin{document}

\title{Experimental realization of stimulated Raman shortcut-to-adiabatic passage with cold atoms}

\author{Yan-Xiong Du$^{1}$, Zhen-Tao Liang$^{1}$, Yi-Chao Li$^{2}$, Xian-Xian Yue$^{1}$, Qing-Xian Lv$^{1}$, Wei Huang$^{1}$, Xi Chen$^{2,}$
\footnote{Correspondence and requests for materials should be
addressed to X. C. (email:xchen@shu.edu.cn), H. Y. (email:
yanhui@scnu.edu.cn) or S. -L. Z. (email: slzhu@nju.edu.cn).}, Hui
Yan$^{1,~*}$, Shi-Liang Zhu$^{3,1,4,~*}$}

\affiliation{$^1$Guangdong Provincial Key Laboratory of Quantum Engineering and Quantum
Materials, SPTE, South China Normal University, Guangzhou 510006, China\\
$^2$Department of Physics, Shanghai University, Shanghai 200444 , China\\
$^3$National Laboratory of Solid State Microstructures, School of
Physics, Nanjing University, Nanjing 210093, China\\
$^4$Synergetic Innovation Center of Quantum Information and
Quantum Physics, University of Science and Technology of China,
Hefei 230026, China}

%\date{\today}

\begin{abstract}
Accurate control of a quantum  system is a fundamental requirement
in many areas of modern science ranging from quantum information
processing to high-precision measurements. A significantly
important goal in quantum control is to prepare a desired state as
fast as possible with sufficiently high fidelity allowed by
available resources and experimental constraints. Stimulated Raman
adiabatic passage (STIRAP) is a robust way to realize
high-fidelity state transfer but it requires a sufficiently long
operation time to satisfy the adiabatic criteria. We here
theoretically propose and then experimentally demonstrate a
shortcut-to-adiabatic protocol to speed up the STIRAP. By
modifying the shapes of the Raman pulses, we experimentally
realize a fast and high-fidelity stimulated Raman
shortcut-to-adiabatic passage that is robust against control
parameter variations.
%, such as the shape distortions and amplitude shifts of the Raman pulses.  Therefore,
The all-optical, robust, and fast protocol demonstrated here
provides an efficient and practical way to control quantum
systems.
\end{abstract}

\maketitle

Coherent control of the quantum state is an essential task in
various areas of physics, such as high-precision measurement
\cite{kasevivh,kotru}, coherent manipulation of atom and molecular
systems \cite{rice-1,kral} and quantum information
\cite{farhi,Monroe}. In most applications, the basic requirement
of coherent control is to reach a given target state with high
fidelity as fast as possible. Many schemes have been developed for
this purpose, including the adiabatic passage technique, which
drives the system along its eigenstate
\cite{vitanov,rangelov,torosov,kovachy}. One of attractive
property of this technique is that the resulting evolution is
robust against control parameter variations when the adiabatic
condition is fully satisfied.  However, the adiabatic passage
techniques such as the two-level adiabatic passage \cite{kovachy},
three-level stimulated Raman adiabatic passage (STIRAP)
\cite{bergmann}, and their variants are time consuming to realize,
which limits their applications in some fast dephasing quantum
systems. To overcome this shortcoming, several protocols within
the framework of the so-called ``shortcut-to-adiabaticity''
\cite{Torrontegui} have been proposed to speed up the ``slow"
adiabatic passage: for instance, counter-diabatic driving
(equivalently, the transitionless quantum algorithm)
\cite{demirplak,demirplaka,berry1,berry}. Very recently, the
acceleration of the adiabatic passage has been demonstrated
experimentally in two-level systems: an energy-level anticrossing
for a Bose-Einstein condensate loaded into an accelerated optical
lattice \cite{bason} and the electron spin of a single
nitrogen-vacancy center in diamond \cite{zhang}.

The STIRAP based on the two-photon stimulated Raman transition has
several advantages. First, lasers can be focused on a single site
in an optical lattice or on a single ion in a linear ion trap,
which guarantees individual addressability
\cite{wang,isenhower,tan}. Second, the STIRAP can couple two
states that can't be directly coupled, such as transferring
population between two atomic states with the same parity (which
can't be directly coupled via electric dipole transition)
\cite{rydberg}, or transferring the atomic state to the molecular
state \cite{rice-1}. Furthermore, with large single-photon
detuning, double coherent adiabatic passages exist
\cite{klein,kleina,du}, which guarantees the capacity for state
transfer between arbitrary states \cite{Lacour,du,huang}.
Interestingly, several theoretical protocols have been proposed to
speed up the STIRAP by adding an additional microwave field in
various atom and molecular systems
\cite{unanyan,chen,Giannelli,rice}. However, the transfer fidelity
will depend on the phase differences among the microwave field,
the Stokes and pumping laser pulses for the STIRAP, which are
difficult to lock. Furthermore, the combination of the microwave
field and Raman lasers makes it difficult to feature the
individual addressability of the operation. Therefore, speeding up
the STIRAP has not yet been experimentally demonstrated.

Motivated by the goal of a robust, fast, addressable, arbitrary
state transfer protocol, we propose a feasible scheme to speed up
STIRAP by modifying the shapes of two Raman pulses. We utilize the
counter-diabatic driving along with unitary transformation, one of
the shortcut techniques to realize adiabatic passages. We then
experimentally demonstrate the proposed stimulated Raman
shortcut-to-adiabatic passage (STIRSAP) protocol in a large
single-photon detuning three-level $\Lambda$ system with a cold
atomic ensemble. The passage's robustness against parameter
variation is confirmed in our experiments. Fast, robust,
individually addressable, and arbitrarily  transferable between
states, the quantum state control protocol demonstrated
here is useful for practical applications.\\

\begin{figure*}[ptb]
\begin{center}
\includegraphics[width=12cm,]{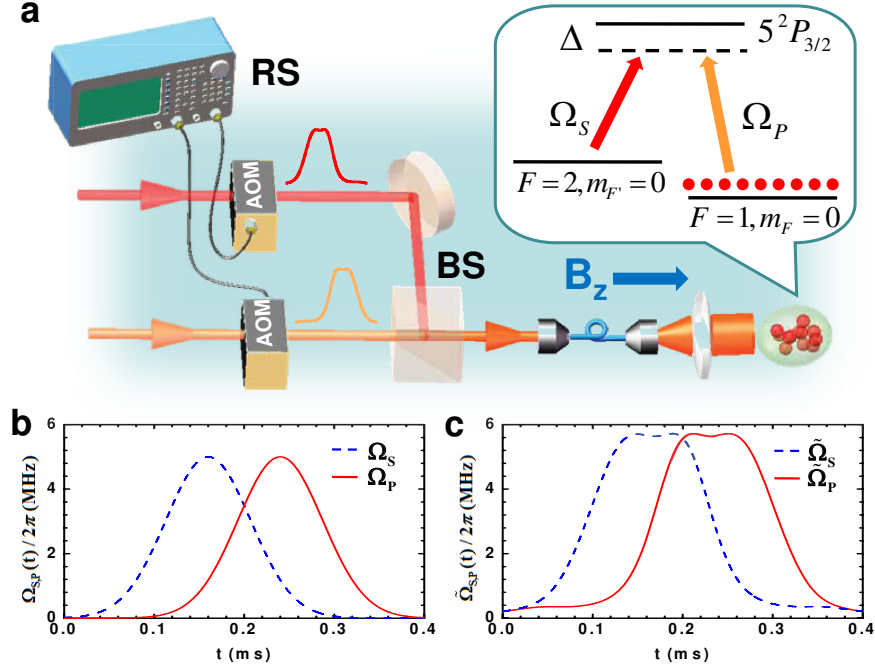} \caption{\label{fig:set up}
\textbf{Experimental scheme.} (a) Experimental setup. The
laser-atom coupling scheme of the three-level $\Lambda$ system is
shown in the upper panel. A magnetic field $B_\textrm{z}$ is used
to split the  Zeeman sublevels. Two Raman laser fields (Stokes
$\Omega_S (t)$ and pumping $\Omega_P (t)$) are combined by a beam
splitter (BS) and then sent to interact with the cold atoms. The
shapes of the Raman lasers are modulated by two acousto-optic
modulators (AOMs) driven by a radio source (RS).
%The single-photon detuning $\Delta$ between the Raman lasers and the
%excited state $5 ^2$P$_{3/2}$ of $^{87}$Rb is about 2.5 GHz.
(b) Original Raman laser pulses for STIRAP. (c) Modified Raman
laser pulses for STIRSAP.}
\end{center}
\end{figure*}

\noindent\textbf{\textsf{Results}}\\
\noindent\textbf{STIRAP and STIRSAP protocols.} We consider a cold
$^{87}$Rb atom ensemble (see the Methods) whose internal energy
states $|1\rangle$ ($|2\rangle$) and $|3\rangle$ are coupled by
pumping pulse $\Omega_P (t)$ [Stokes pulse $\Omega_S (t)$], as
shown in Fig. 1a. Two ground states $|F=1, m_F=0\rangle
=|1\rangle$, $|F=2, m_{F}=0\rangle=|2\rangle$ and one excited
state $5^2$P$_{3/2}$ ($=|3\rangle$) are selected as a typical
three-level $\Lambda$ system. Under the conditions of
rotating-wave approximation and two-photon detuning $\delta=0$,
the interaction Hamiltonian of the system in the basis of
$\{|1\rangle, |2\rangle, |3\rangle\}$ is given as
\begin{equation}
\label{H0} H_\Lambda(t)=\frac{\hbar}{2}\left(\begin{array}{ccc}
0&0&\Omega_{P}(t)e^{i\varphi_L}\\
0&0 & \Omega_{S}(t)\\
\Omega_{P}(t)e^{-i\varphi_L}&\Omega_{S}(t)&2\Delta
\end{array}\right),
\end{equation}
where $\Delta$ is the single-photon detuning and  $\varphi_L$ is
the phase difference between Stokes and pumping lasers and has
been locked to a fixed value in our experiment. In the large
detuning condition $\Delta \gg
\sqrt{\Omega_P^2(t)+\Omega_S^2(t)}$, the three dressed states of
the Hamiltonian (\ref{H0}) can be described as
$|D\rangle=\cos\theta |1\rangle-\sin\theta \exp(-i\varphi_L)
|2\rangle$, $|B_1\rangle\simeq\sin\theta\exp(i\varphi_L)
|1\rangle+\cos\theta|2\rangle$, and $|B_2\rangle\simeq|3\rangle$,
where mixing angle $\theta=\arctan [\Omega_P (t)/\Omega_S (t)]$
\cite{du,Zhu}. In the usual STIRAP protocol, the Stokes and
pumping laser pulses are partially overlapping Gaussian shapes
\cite{bergmann}. If the adiabatic condition $ T \gg T_\pi$ is
fulfilled, where $T$ is the operation time and
$T_\pi=2\pi\Delta/(\Omega_P\Omega_S)$ with $\Omega_P$ and
$\Omega_S$ being the respective peaks of the pulses $\Omega_P(t)$
and $\Omega_S (t)$, a high-fidelity coherent population transfer
from one specific superposition state of $|1\rangle$ and
$|2\rangle$ to another can be realized through adiabatic evolution
of the dressed states $|D\rangle$ and $|B_1\rangle$. This protocol
is the double coherent STIRAP \cite{du} we used in our
experiments.

\begin{figure*}[ptb]
\begin{center}
\includegraphics[width=12cm,]{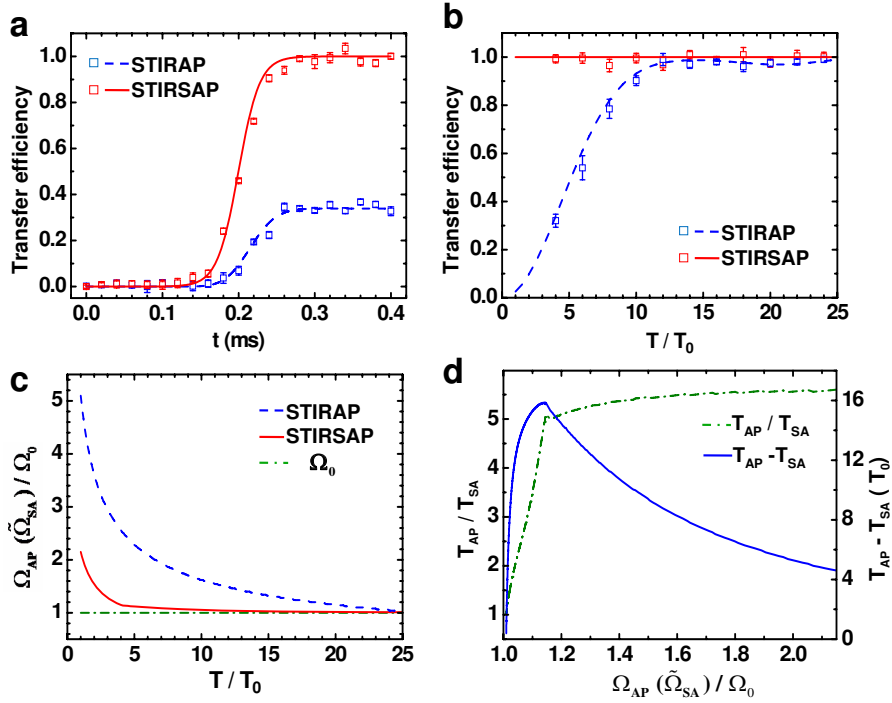}
\caption{
%(color online)
\textbf{The speedup results.} (a) Experimental (squares) and
theoretical (lines) results of population transfer dynamics. (b)
Transfer efficiency versus operation time $T$. (a) and (b) are
driven by the Raman pulses plotted in figure 1b (blue dashed line
and blue squares) and figure 1c (red dashed line and red squares).
The transfer efficiency of STIRAP can only reach $36\%$; in
contrast, that for STIRSAP can approach $100\%$. The data points
in (a) and (b) are averaged over five measurements, each with the
error bars depicting the standard deviation. (c) Maximum Rabi
frequency $\tilde{\Omega}_{SA}$ of STIRSAP (red solid line) and
$\Omega_{AP}$ of STIRAP (blue dashed line) versus operation time
$T$ with the same fidelity. The original Raman frequency
$\Omega_0$ is plotted as green dotted-dashed line. (d) Comparison
of time $T_{AP}$ of STIRAP and $T_{SA}$ of STIRSAP to achieve the
same 99.4\% efficiency and with equal maximum Rabi frequency
($\Omega_{AP}=\tilde{\Omega}_{SA}$). Ratio $T_{AP}/T_{SA}$ (green
dotted-dashed line) approaches 5.6 as $\Omega_{AP}$
($\tilde{\Omega}_{SA})$ increases, which indicates the maximum
acceleration we can obtain. Difference $T_{AP}-T_{SA}$ is plotted
in blue solid line  where the maximum shows that the optimal
STIRSAP is reached at $\tilde{\Omega}_{SA}/\Omega_0=1.14$.}
\end{center}
\end{figure*}

To release the critical requirement $ T \gg T_\pi$ but still
maintain the high-fidelity, one can adopt the  shortcut approach
to adiabatic passage \cite{demirplak,demirplaka,berry}.
% initially proposed by Berry\cite{berry}.
Under the large detuning condition,  the population in excited
state $|3\rangle$ can be  adiabatically eliminated. The
Hamiltonian (\ref{H0}) can then be reduced into an effective
two-level system on the basis $\{|1\rangle,|2\rangle\}$, and the
Hamiltonian is given by
\begin{equation}
\label{H_eff} H_{0}(t)=-\frac{\hbar}{2}\left(\begin{array}{cc}
\Delta_{eff} & \Omega_{eff}e^{i\varphi_L}\\
\Omega_{eff}e^{-i\varphi_L} & -\Delta_{eff}\\
\end{array}\right),
\end{equation}
 where the effective detuning $\Delta_{eff}=[\Omega^2_{P}(t)-\Omega^2_{S}(t)]/({4\Delta})$ and the effective Rabi frequency
 $\Omega_{eff}={\Omega_{P}(t)\Omega_{S}(t)}/({2\Delta}$).
 According to the standard shortcut approach  to adiabatic passage, the diabatic transition
can be eliminated by adding an appropriate auxiliary
counter-diabatic term $H_{cd} (t)$ defined in the Methods
\cite{berry,Torrontegui}. In our system, this auxiliary term
$H_{cd} (t)$ can be realized by adding a microwave field to couple
the levels $\{|1\rangle$ and $|2\rangle\}$ \cite{Giannelli,chen};
however, the aforementioned drawbacks of this method still need to
be overcome.
%, such as the phase between the microwave and the Raman lasers should be locked, and it is quilt complicated.

In the Methods section, we describe a feasible approach to realize
the shortcut method to adiabatic passage. We find that
high-fidelity STIRSAP can be achieved if the shapes of the Raman
pulses are replaced by
\begin{equation}
\begin{split}
\label{Omega}
\tilde{\Omega}_P(t)=\sqrt{2\Delta(\sqrt{\tilde{\Delta}_{eff}^2(t)+\tilde{\Omega}_{eff}^2(t)}+\tilde{\Delta}_{eff}(t))},\\
\tilde{\Omega}_S(t)=\sqrt{2\Delta(\sqrt{\tilde{\Delta}_{eff}^2(t)+\tilde{\Omega}_{eff}^2(t)}-\tilde{\Delta}_{eff}(t))},
\end{split}
\end{equation}
where $\tilde{\Delta}_{eff}(t)$, $\tilde{\Omega}_{eff}(t)$ are
respectively the modified effective detuning and Rabi frequency as
defined in the Methods section. The modified Raman pulses still
satisfy the large detuning condition. With appropriate choices of
the parameters $\tilde{\Omega}_P(t)$ and $\tilde{\Omega}_S(t)$,
the system is effectively equivalent to that of adding a
supplementary counter-diabatic term $H_{cd} (t)$
\cite{bason,saraPRL}. The system will thus evolve along its
eigenstate of the Hamiltonian $H_{0} (t)$ up to the phase factor
for any choice of the protocol parameters, even with very small
values of Stokes and pumping fields and within an arbitrarily
short operation time $T$. According to Eq. (\ref{Omega}), given
the original Stokes and pumping pulses with the Gaussian-beam
shape shown in Fig. 1b,
%used in $H_0(t)$,
the modified Stokes and pumping pulses required for
STIRSAP can be obtained as shown in Fig. 1c.
\\

\begin{figure*}[tpb]
\begin{center}
\includegraphics[width=17cm,,height=5cm]{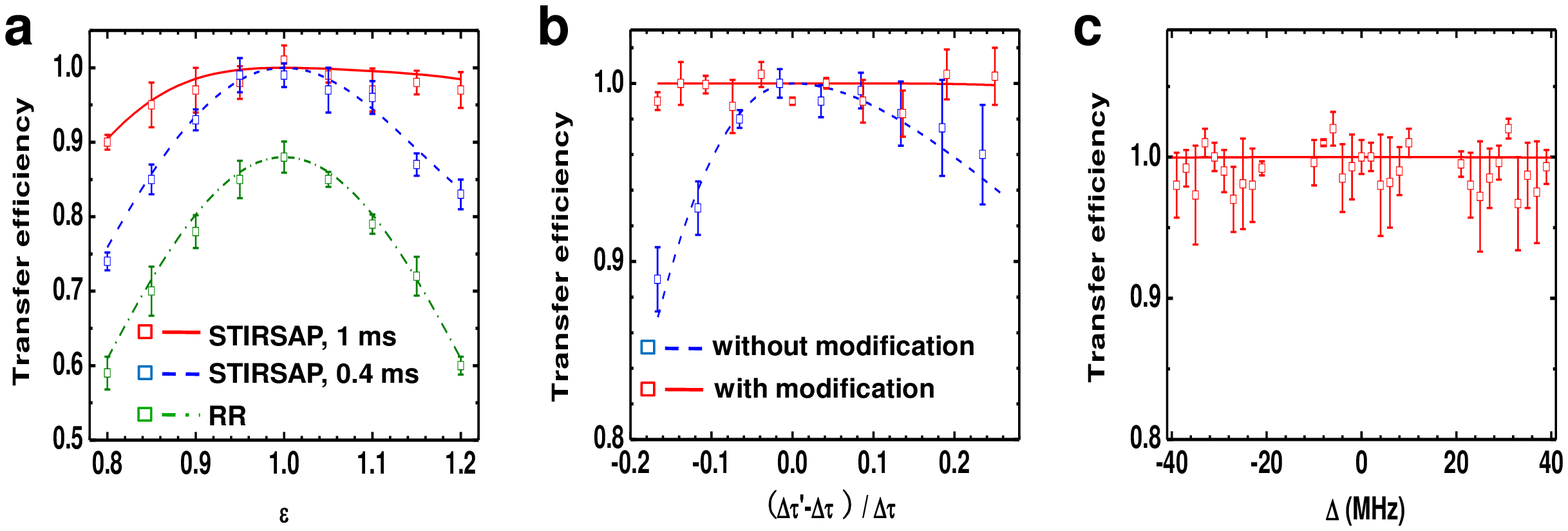}
\caption{
%(color online)
\textbf{Transfer efficiencies of STIRSAP versus imperfections.}
There are three types of imperfections being discussed: (a)
Variation in the peak of Rabi frequency characterized with
parameter $\varepsilon$. Theoretical curves (solid and dashed
lines) and experimental data (squires) correspond with operation
time $T=1$ (0.4) ms. Theoretical curve (green dotted-dashed line )
and experimental data (green squares) represent the result of
resonant Raman (RR) $\pi$ pulse. (b) Variation in separation time.
Blue squares and dashed lines correspond to variations in
$\Delta\tau'$ without pulses shape modification; red squares and
solid lines correspond to variations with modification. (c)
Variation in single-photon detuning $\Delta$. The data points are
averaged over five measurements, each with the error bars
depicting the standard deviation.}
\end{center}
\end{figure*}

\noindent \textbf{Dynamics and characteristics.} We now compare the
performance of the above STIRAP and STIRSAP protocols. In our
experiment, the Stokes pulse
$\Omega_S(t)=\Omega_S\exp{[-(t-T/2+\Delta\tau)^2/\sigma^2]}$ and
pumping pulse
$\Omega_P(t)=\Omega_P\exp{[-(t-T/2-\Delta\tau)^2/\sigma^2}]$, where
2$\sigma=T/3$ is the full width at half maximum of the pulse, and
$\Delta\tau=T/10$ is the separation time between the two pulses. We
first compare the population transfer dynamics with Raman pulses as
shown in Fig. 1b and 1c. The original parameters of STIRAP are set
to be $\Delta \sim 2\pi \times 2.5$ GHz, $\Omega_P=\Omega_S=2\pi
\times 5$ MHz, and hereafter we denote $\Omega_0 \equiv 2\pi \times
5$ MHz and the corresponding $\pi$ pulse time $T_0\equiv 2\pi
\Delta/\Omega_0^2=0.1$ ms. Experimental data (blue and red squares)
and theoretical results (dashed and solid lines) are shown together
in Fig. 2a. Here the operation time $T=0.4$ ms, which fails to
fulfill the adiabatic criteria. As shown in Fig. 2a, the final
transfer efficiency of STIRAP only reaches 36\% (blue dashed line).
As for the STIRSAP Raman pulses implemented by replacing
$\Omega_{P,S} (t)$ with $\tilde{\Omega}_{P,S} (t)$ in Eq.
(\ref{Omega}), the transfer efficiency (the red solid line) can
reach 100\% since the diabatic transition has been eliminated by
effectively adding the Hamiltonian $H_{cd} (t)$. The peak transfer
efficiencies of STIRSAP are observed with a two-photon detuning
$\delta=-7$ kHz due to ac-Stark shift.  The ac-Stark shift can be
viewed as a perturbation in our case since it is small compared to
$\Omega_{S,P}$ and the two-photon bandwidth ( $\sim 20$ kHz)
\cite{du}. The experimental and theoretical results fit very well
with each other. This result clearly shows the remarkable feature of
the STIRSAP protocol.

To further characterize the performance of STIRAP and STIRSAP, we
plot the transfer efficiencies of them as a function of operation
time $T$ in Fig. 2b for a fixed $\Omega_{P,S}=\Omega_0$. With
STIRAP, the transfer efficiency approaches $100\%$ when the
operation time is longer than $25T_0$, where the adiabatic
condition is fully satisfied \cite{bergmann}; however, the
efficiency (blue-dashed line) will decrease along with the
decreasing of $T$. In particular, it decreases quickly when $T<10
T_0$. Remarkably, it is shown in theoretical calculation that the
transfer efficiency of STIRSAP (red solid line) can keep constant
for any operation time $T$ since the diabatic transition has been
eliminated by effectively adding the $H_{cd}$ term through
modifying the shape of the pulses accordingly. We confirm the
theoretical result with the experimental data for $T\geq 4 T_0$,
where the peak of $\tilde{\Omega}_{P,S} (t)$ is around
$1.14\Omega_0$ for $T=4 T_0$.

In principle, both STIRAP and STIRSAP can be sped up to a fixed
operation time with fidelity higher than certain value if the
peaks of Raman pulses are sufficiently large; however, the
resources required are different. With STIRAP, we denote the peak
of $\Omega_{S,P} (t)$ as $\Omega_{AP}$. Because the characterized
time for adiabatic evolution $T_\pi=2\pi\Delta/\Omega_{AP}^2$
decreases with increasing  $\Omega_{AP}$,  the operation time can
decrease even for a fixed fidelity. By contrast, as shown in Fig.
2b,  the operation time for STIRSAP can be  arbitrarily small by
suitably choosing the peak $\tilde{\Omega}_{SA}$ of the modified
Raman pulses $\tilde{\Omega}_{P,S} (t)$. To address the resources
required for the speedups, we plot in Fig. 2c the peaks
$\Omega_{AP}$ (blue dashed line) and $\tilde{\Omega}_{SA}$ (red
solid line) required for operation time $T$ with fidelity no less
than 99.4$\%$. It is clear that peak $\tilde{\Omega}_{SA}$ is much
smaller than  $\Omega_{AP}$ for the same operation time with the
same high fidelity. This reveals that for the same time $T$ and
same fidelity, the resources required for STIRSAP is less than
that for STIRAP.

To further compare the performance of STIRAP and STIRSAP, we test
the maximum capability of speedup that we could obtain for equal
maximum Rabi frequencies, i.e., $\Omega_{AP}=\tilde{\Omega}_{SA}$.
We theoretically calculate the time $T_{AP}$ of STIRAP to achieve
the same high fidelity ($99.4\%$) transfer by sweeping
$\Omega_{AP}$  and then compare $T_{AP}$ with the operation time
$T_{SA}$ for STIRSAP by sweeping $\tilde{\Omega}_{SA}$. As shown
by the green dashed line in Fig. 2d, for the initial Rabi
frequency of $\tilde{\Omega}_{SA}=\Omega_0$, which corresponds to
a long operation time $T_{SA}$, the auxiliary Rabi frequency
$\Omega_a$ is small, resulting in only a slight improvement in
$T_{SA}$ (see the time-derivation term in Eq. (5) in Methods).
However, if we slightly increase $\tilde{\Omega}_{SA}$, $\Omega_a$
increases, while the ratio $T_{AP}/T_{SA}$ quickly increases. The
ratio is finally stabilized at $5.6$, which means that STIRSAP can
achieve a speedup $5.6$ times that of STIRAP for a fixed
$\Omega_0$. Although the maximum speedup is achieved when
$\tilde{\Omega}_{SA}$ is larger than 2$\Omega_0$, an optimal
speedup can be achieved by increasing a moderate factor in
$\Omega_0$. We also plot the difference $T_{AP}-T_{SA}$ (in unit
of $T_0$) as shown in Fig. 2d (solid blue line) which reaches its
maximum when $\tilde{\Omega}_{SA}\approx1.14\Omega_0$.
\\

\noindent\textbf{Robustness against imperfection.} We now test the
stability of the STIRSAP protocol with respect to control
parameter variations. To this end, we experimentally measure and
theoretically calculate the transfer efficiency by varying one of
the protocol parameters in Hamiltonian (\ref{H0}) (i.e., the
amplitudes $\tilde\Omega_{SA}$ and relative time delay $\Delta
\tau$ of the Stokes and pumping pulses, and single-photon detuning
$\Delta$) while keeping all other parameters unchanged.

The amplitude of the Raman pulses for each atom in our system is
slightly different since there is a space distribution of laser
power around $\pm$5$\%$ on the atomic cloud. Here we artificially
modify the amplitudes of the Raman pulses as
$\Omega'_{RR}=\varepsilon\Omega_{RR}$ and
$\tilde{\Omega}'_{SA}=\varepsilon\tilde{\Omega}_{SA}$,
$\epsilon\in[0.8,1.2]$ (where $RR$ represents resonant Rabi
pulses) to simulate the amplitude variation. Figure 3a shows the
experimental data (squares) and theoretical  results (lines)  of
the transfer efficiencies as a function of the deviation
$\varepsilon$ for the resonant Raman $\pi$ pulse (green squares
and dotted-dashed line ), STIRSAP with $T=0.4$ ms (blue squares
and dashed line) and STIRSAP with $T=1$ ms (red squares and solid
line).  As shown in Fig. 3a, the resonant Raman $\pi$ pulse is
very sensitive to the amplitude variation of Rabi frequencies, and
the maximum transfer efficiency is less than $90\%$ due to the
intensity space distribution of laser fields. Remarkably, the
STIRSAP is less sensitive to the change of $\tilde{\Omega}'_{SA}$,
since the system adiabatically evolves along the eigenstate of
Hamiltonian $H_0$, which depends only on the ratio of the Stokes
and pumping fields. The robustness will be improved if we extend
$T=0.4$ ms to $T=1$ ms, because it will be easier for the system
to follow the changes of the ratio of the Stokes and pumping
fields.

The transfer efficiencies as a function of the separation time are
plotted in Fig. 3b. We first measure the transfer efficiency with
fixed pulses shapes versus different separation times
$\Delta\tau'$. The pulses of STIRSAP are generated with parameters
$\Delta\tau=T/10$ and $T=0.4$ ms. The real separation time
$\Delta\tau'$ in our system is achieved by triggering the radio
resource with a delay time at a range about $\pm 20\%$ in
$\Delta\tau$. We observe the largest  $10\%$ reduction in
efficiency  as shown by the blue squares in Fig. 3b, which accords
with the theoretical simulation (blue dashed line).  We then
measure the transfer efficiency with variable pulse shapes versus
different separation times. Here the Raman pulses  we use for
every separation time are calculated for the STIRSAP according to
each specific separation time. Under this condition, the transfer
efficiency can be kept to almost 1 as shown by the red curves and
squires in Fig. 3b.

We further test the sensitivity of the STIRSAP protocol to the
variation of the single-photon detuning $\Delta$ in Hamiltonian.
The detuning $\Delta$ can be changed in the range of $\pm40$ MHz
in our experiment. The frequency adjustment is implemented by
changing the radio frequencies of AOMs and the locking points of
the pump laser. There are three locking points
($F=2\longleftrightarrow F'=2$, $F=2\longleftrightarrow F'=3$, and
the crossover peak between them) in our setup, and the radio
frequencies of AOMs can be continuously varied $\pm10$ MHz around
each locking point. Although a specific  single-photon detuning
$\Delta$ is needed in the calculation of the STIRSAP protocol (see
Eq. (3)), as shown in Fig. 3c, the transfer efficiency keeps
constant  as frequency changes, which indicates that STIRSAP will
not suffer from the deviation of the detuning $\Delta$, since the
variation of $\Delta$ is less than 1 MHz in the experiments.

As discussed above, in the region where the relative imperfection
is less than $5\%$, STIRSAP with $T=0.4$ ms can maintain a
fidelity higher  than $98\%$, which shows a good robust feature
for potential applications in quantum manipulation.
\\

\noindent\textbf{Double coherent passages and multiple cyclic
operation.} So far, we have demonstrated that the STIRSAP protocol
is fast, robust, and has a high fidelity. As a further proof of
its fast and high-fidelity features, we apply STIRSAP pulses at
the  maximum speedup point ($T=0.4$ ms for $\Omega_{0}$) five
times to realize back-and-forth operations in our system. It is
noted that the total operation time is limited to  3 ms in our
system, mainly due to the expansion of the atomic cloud. For the
large single-photon detuning $\Lambda$ system, two coherent
passages exit. Thus the state can be cycled back-and-forth with
the same order of Raman pulses. As shown in Fig. 4a, we first pump
all the atoms to one of the ground states ($|1\rangle$) and then
repeat the STIRSAP pulse five times. The system will evolve along
one eigenstate and then another one. The final population transfer
efficiency to the other ground state ($|2\rangle$) is $(95 \pm 4)
\%$ averaged over five measured data sets, which indicates an
average efficiency of 99(6)$\%$.

More interestingly, the STIRSAP protocol with double coherent
passages demonstrated here can also be used to drive the
superposition state, which is impossible in ordinary STIRAP with
zero detuning. As for an example, we experimentally realize a
$\sigma_x$ gate between the initial superposition state
$|\psi_0\rangle=\sqrt{0.3}|1\rangle+e^{i\phi_0}\sqrt{0.7}|2\rangle$
and the final state
$|\psi_0\rangle=e^{i\phi_0}\sqrt{0.7}|1\rangle+\sqrt{0.3}|2\rangle$
with $\phi_0$ an irrelevant phase. The data driven back-and-forth
for five times are shown in Fig. 4b. Comparing with the ideal
population $0.7$ in state $|1\rangle$, the final population
measured after five $\sigma_x$ operations is $(68 \pm 4)\%$, which
indicates a total transfer efficiency of $96(8)\%$ and an average
efficiency of 99(5)$\%$. Note that those multiple cycle operations
in Fig. 4a,b can not be implemented by STIRAP in our system due to
the time limit from the expansion of the atomic cloud. The results
thus show remarkable advantages of STIRSAP in some quantum systems
with short coherent time.
\\

\begin{figure}[tpb]
\begin{center}
\includegraphics[width=9cm,]{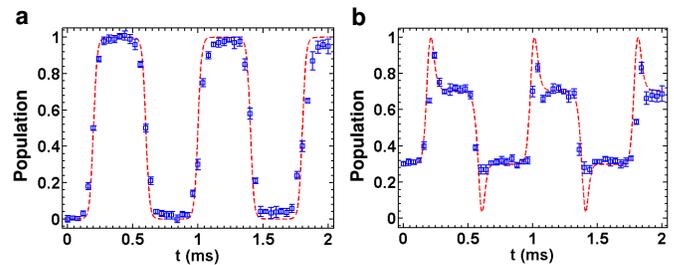}
\caption{
%(color online).
\textbf{ Experimental realization of multiple cycle operations.}
(a) with initial  state $|1\rangle$. (b) with initial state
$|\psi_0\rangle=\sqrt{0.3}|1\rangle+e^{i\phi_0}\sqrt{0.7}|2\rangle$.
For both (a) and (b), blue squares are the experimental data, and
red dashed lines are theoretical results under ideal conditions.
The data points are averaged over five measurements, each with the
error bars depicting the standard deviation. }
\end{center}
\end{figure}

\noindent\textsf{\textbf{Conclusion}}\\
In summary, we have theoretically proposed and experimentally
demonstrated an useful protocol to speed up conventional ``slow"
STIRAP in a large single-photon detuning three-level system
through transitionless passage. The STIRSAP demonstrated here is
faster than STIRAP and more robust as compared to resonant Raman
$\pi$ pulses. Furthermore, the existence of double coherent
passages provides a feasible way to control arbitrary quantum
states. Fast, high in fidelity,  and robust against control
parameter variations, the STIRSAP protocol is promising for
practical applications in quantum control, quantum information
processing, and even chemical interaction control.
\\

\noindent\textsf{\textbf{Methods}}\\
\noindent\textbf{Cold atomic ensemble controlled by Raman lasers.}
Our experimental  system shown in Fig. 1a is similar to the one
described in our previous work \cite{du}. The $^{87}Rb$ atoms are
trapped by a magneto-optical trap. Two Raman lasers (Stokes and
pumping lasers) respectively couple two ground states
($|1\rangle$, $|2\rangle$) with the excited state ($|3\rangle$).
The Raman lasers are set to be two-photon resonance ($\delta=0$)
and large single-photon detuning ($\Delta \sim 2\pi \times 2.5$
GHz) from the excited state. The frequency of the Stokes laser is
further locked to the pumping laser with a stable beating
frequency (bandwidth is less than 0.1 kHz) through optical
phase-locked loop technique. The shapes of Raman pulses are
controlled by two acousto-optic modulators (AOMs in Fig. 1a),
which are driven by a radio source (Rigol, DG4162). The radio
source has a frequency stability smaller than 2ppm and a maximum
frequency output of 160MHz.

With a bias field $B_\textrm{z}$ about 0.1 Gauss, two-photon Raman
transition between magnetic sublevels of $|F=1\rangle$  and
$|F=2\rangle$ is split by 140 kHz, which allows us to selectively
transfer population between $|F=1, m_F=0\rangle$ and $|F=2,
m_{F'}=0\rangle$. Population is measured with the fluorescence
collected by a photodiode. To eliminate the total population
fluctuation, the populations of $|F=1, m_F=0\rangle$ and $|F=2,
m_{F'}=0\rangle$ are measured simultaneously in the experiments
for normalization.
\\

\noindent\textbf{Detailed STIRSAP method.} Under the large
detuning condition, the three-level $\Lambda$ system reduces to
an effective two-level system described by the Hamiltonian
(\ref{H_eff}).
 According to the
theory of shortcut-to-adiabatic passage, the diabatic transition
can be eliminated by adding a counter-diabatic term given as
$H_{cd}(t)=i\hbar\sum(|\partial_t\lambda_n\rangle\langle\lambda_n|-\langle\lambda_n|\partial_t\lambda_n\rangle|\lambda_n\rangle\langle\lambda_n|)$
\cite{berry,chen}, which will lead the system evolution along the
eigenstate $|\lambda_n\rangle$ $(=\{ |D\rangle,\ |B_1\rangle \}$
here ) for any time $T$. For our system, the counter-diabatic term
can be realized by adding a microwave field to couple the levels
$|1\rangle$ and $|2\rangle$ \cite{Giannelli,chen}.  Given this,
the counter-diabatic term $H_{cd}$ should be given by
\begin{equation}
H_{cd}(t)=\frac{\hbar}{2}\left(\begin{array}{ccc}
0 & \Omega_a(t)e^{i\varphi_a}\\
\Omega_a(t)e^{-i\varphi_a} & 0\\
\end{array}\right),
\end{equation}
where
\begin{equation} \label{Omega_a}
\Omega_a(t)=\frac{2[\dot{\Omega}_{P}(t)\Omega_{S}(t)-\Omega_{P}(t)\dot{\Omega}_{S}(t)]}{\Omega^2_{P}(t)+\Omega^2_{S}(t)}
\end{equation}
represents the Rabi frequency of the auxiliary-driving field and
its phase $\varphi_a=\varphi_L+\pi/2$. The phase relation requires
one to lock the phase between the microwave field and the Raman
lasers, which is quite complicated.

To overcome these drawbacks, we develop a much simpler approach to
realize the shortcut method to adiabatic passage. We note that
$H_{cd}$ can be absorbed into the variation of the original field
to form a total Hamiltonian, $H(t)=H_{0}(t)+H_{cd}(t)$, given by
\begin{equation}
\begin{aligned}
H(t)=-\frac{\hbar}{2}\left(\begin{array}{ccc}
\Delta_{eff} & \sqrt{\Omega^2_{eff}+\Omega^2_a}e^{-i\gamma (t)}\\
\sqrt{\Omega^2_{eff}+\Omega^2_a}e^{i\gamma (t)} & -\Delta_{eff}\\
\end{array}\right),
\label{Ht}
\end{aligned}
\end{equation}
where $\gamma (t)= \phi(t)+\varphi_L$ with
$\phi(t)=\arctan(\Omega_a(t)/\Omega_{eff}(t))$.
%Since $\phi_L$ is
%a common phase in Eq. (3) and Eq.(9),  we set $\phi_L=0$ which
%will not influence the discussion below.
It implies that the additional microwave field to achieve $H_{cd}$
is not necessary. We may simply modify  both the phase and the
amplitude of the Raman lasers to effectively add the $H_{cd}$ term
and thus realize the shortcut-to-adiabatic passage protocol.
Moreover, we further show that the precise  control of the
time-dependent phase $\gamma (t)$, which is still complicated, can
be released. To this end, we apply the unitary transformation
\cite{berry1,bason,saraPRL}
\begin{equation}
U(t)=\left(\begin{array}{ccc}
e^{-i\gamma(t)}/2 & 0\\
0 & e^{i\gamma(t)/2}\\
\end{array}\right),
\end{equation}
which amounts a rotation around the $Z$ axis by $\gamma$ and
eliminates the $\sigma_y$ term in the Hamiltonian (\ref{Ht}).
After the transformation, we obtain an equivalent Hamiltonian with
Eq. (\ref{Ht}), $\tilde{H}(t)=U^\dagger HU-i\hbar U^\dagger
\dot{U}$, that is,
\begin{equation}
\begin{aligned}
\tilde{H}(t)
=-\frac{\hbar}{2}\left(\begin{array}{ccc}
\tilde{\Delta}_{eff}(t) & \tilde{\Omega}_{eff}(t)\\
\tilde{\Omega}_{eff}(t) &-\tilde{\Delta}_{eff}(t)\\
\end{array}\right),
\label{H'}
\end{aligned}
\end{equation}
where the modified effective detuning
$\tilde{\Delta}_{eff}(t)=\Delta_{eff}(t)+\dot{\phi}$ and effective
Rabi frequency
$\tilde{\Omega}_{eff}(t)=\sqrt{\Omega^2_{eff}(t)+\Omega^2_a (t)}$.
In the derivation, $\dot{\varphi}_L=0$ is used. The wave function
$|\tilde{\Psi}(t)\rangle$ related to the Hamiltonian $\tilde{H}(t)$
is $|\tilde{\Psi}(t)\rangle=U|{\Psi}(t)\rangle$, where
$|{\Psi}(t)\rangle$ is the wave function related to the Hamiltonian
$H(t)$ in Eq. (\ref{Ht}). Since the unitary transformation $U(t)$ is
diagonal and the elements are just phase factors, population
measured in the basis $\{|1\rangle,|2\rangle\}$ should be the same
for both $|\tilde{\Psi}\rangle$ and $|{\Psi}\rangle$.

An interesting result implied in Eq. (\ref{H'}) to further
simplify the experimental protocol, which will be proven in the
next section, is that we can realize shortcut-to-adiabatic passage
by replacing $\Omega_S(t)$ and $\Omega_P(t)$ in Hamiltonian
(\ref{H0}) with modified Raman pulses $\tilde{\Omega}_S(t)$,
$\tilde{\Omega}_P(t)$. By solving the following equations
\begin{equation}
\begin{split}
\tilde{\Delta}_{eff}(t)=\frac{\tilde{\Omega}^2_P(t)-\tilde{\Omega}^2_S(t)}{4\Delta},\\
\tilde{\Omega}_{eff}(t)=\frac{\tilde{\Omega}_P(t)\tilde{\Omega}_S(t)}{2\Delta},
\end{split}
\end{equation}
we obtain the results of Eq. (\ref{Omega}). Therefore, we can
achieve STIRSAP by replacing the original Raman pulse shapes
$\Omega_{S,P} (t)$ with $\tilde{\Omega}_{S,P} (t)$ as described in
Eq. (\ref{Omega}).

We should point out that, after modifying Raman pulse shapes
$\tilde{\Omega}_{S,P} (t)$, the STIRSAP protocol is robust against
the control parameter variation but is not necessarily optimal.
STIRSAP might be further optimized by using inverse engineering
\cite{ruschhaupt,daems}. Finally, similar STIRSAP protocols can
also be implemented with ordinary single-photon resonant STIRAP of
the three-level system, which can  be reduced to an effective
two-level system due to its intrinsic SU(2) symmetry
\cite{Baksic}.
\\

\noindent\textbf{Dynamics of the three Hamiltonians.} We here prove
that the STIRSAP protocol can be directly achieved by the
realization of Eq. (\ref{H'}). To this end, we compare the dynamics
of the three Hamiltonians $H_0 (t)$, $H (t)$ and $\tilde{H}(t)$. For
any $2\times 2$ Hamiltonian $H^\prime$, we can relate it with an
effective magnetic field $\mathbf{B}^\prime$ by the relation
$H^\prime=\frac{1}{2}\mathbf{\sigma} \cdot \mathbf{B}^\prime$, that
is,
\begin{eqnarray*}
 B_x^\prime &=&H_{12}^\prime+H_{21}^\prime,\\
 B_y^\prime &=&i(H_{12}^\prime-H_{21}^\prime),\\
 B_z^\prime &=&H_{11}^\prime-H_{22}^\prime.
\end{eqnarray*}
The unit vector of the effective magnetic field is defined as
$\hat{\mathbf{B}}^\prime={\mathbf{B}}^\prime/|{\mathbf{B}}^\prime|$.
Replaced $H^\prime$ with  the Hamiltonian $H_0(t)$ in Eq.
(\ref{H_eff}) [the Hamiltonian $H(t)$ in Eq. (\ref{Ht})], we can
obtain such effective magnetic field $\hat{\mathbf{B}}_0$
($\hat{\mathbf{B}}$) for $H_0(t)$ [$H(t)$], and the results are
plotted in Fig. 5, where $\Omega_P=\Omega_S=2\pi \times 5$ MHz,
%for $\mathbf{B}_0$, $\tilde{\Omega}_P$ and $\tilde{\Omega}_S$ are
%obtained from Eq.(3). Other parameters are
$\Delta=2\pi\times 2.5$ GHz, and $T=0.4$ ms.

\begin{figure}[ptb]
\begin{center}
\includegraphics[width=8.2cm]{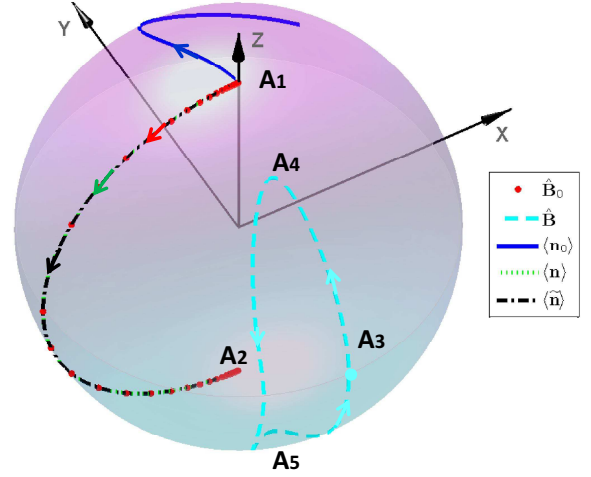}
\caption{\label{Fig5} \textbf{Trajectories of the effective
magnetic fields and the dynamics of the spin polarizations}. The
effective magnetic field $\hat{\mathbf{B}}_{0}$ (solid red dot )
evolves from the north pole $A_1$ to the south pole $A_2$ along
the great circle for the STIRAP protocol. For comparison, the
$\hat{\mathbf{B}}$ (dashed cyan line) for STIRSAP started from
$A_3$ is also shown. Evolution tracks of the initial state
$|1\rangle$ driven by the Hamiltonians $H_{0}$, $H$ and
$\widetilde{H}$, are represented by the spin polarizations
$\langle \mathbf{n}_{0}\rangle$ (solid blue line), $\langle
\mathbf{n}\rangle$ (dotted green line) and $\langle
\widetilde{\mathbf{n}}\rangle$ (dot-dashed black line),
respectively. Since the adiabatic condition is not fully
satisfied, $\langle \mathbf{n}_{0}\rangle$
 doesn't follow $\hat{\mathbf{B}}_{0}$. However,
both $\langle \mathbf{n}\rangle$ and $\langle
\widetilde{\mathbf{n}}\rangle$  evolve exactly along the
trajectory of $\hat{\mathbf{B}}_{0}$, as expected by the STIRSAP
protocol.  The parameters we use to perform numerical simulations
are the same as those in Fig. 2a.}
\end{center}
\end{figure}

Furthermore, we denote $|\Psi_0(t)\rangle$ as the wave function
related to the Schrodinger equation
$i\hbar\partial_t|\Psi_0(t)\rangle= H_0(t)|\Psi_0(t)\rangle$, and
similar denotations for $|\Psi(t)\rangle$ and
$|\tilde{\Psi}(t)\rangle$, then the spin polarizations can be
defined as
\begin{eqnarray*}
\langle \mathbf{n}_0^{x,y,z}(t)\rangle &=&\langle \Psi_0(t)|\sigma_{x,y,z}|\Psi_0(t)\rangle,\\
\langle \mathbf{n}_{x,y,z}(t)\rangle &=&\langle \Psi(t)|\sigma_{x,y,z}|\Psi(t)\rangle,\\
\langle \tilde{\mathbf{n}}_{x,y,z}(t)\rangle &=& \langle
\tilde{\Psi}(t)|\sigma_{x,y,z}|\tilde{\Psi}(t)\rangle.
\end{eqnarray*}
We numerically solve the Schr\"{o}dinger equations for those
Hamiltonians with the initial states given by $|\Psi_0(0)\rangle=
|\Psi(0)\rangle= |\tilde{\Psi}(0)\rangle=|1\rangle$ and the
initial effective magnetic field $\hat{\mathbf{B}}_0(0)$ is along
the $z$ direction. The numerical results of the spin polarizations
are plotted in Fig. 5. If the adiabatic condition is fully filled,
$\langle\mathbf{n}_0 (t)\rangle$ should follow the direction of
$\hat{\mathbf{B}}_0(t)$, but as shown in Fig. 5,
$\langle\mathbf{n}_0 (t)\rangle$ for $T=0.4$ ms does not overlap
$\hat{\mathbf{B}}_0(t)$. However, both
$\langle\mathbf{n}(t)\rangle$ and $\langle
\tilde{\mathbf{n}}(t)\rangle$ follow along the trajectory of
$\hat{\mathbf{B}}_0(t)$. Therefore, rather than following
$\hat{\mathbf{B}}(t)$ or $\hat{\tilde{\mathbf{B}}}(t)$, both
$\langle\mathbf{n}(t)\rangle$ and $\langle
\tilde{\mathbf{n}}(t)\rangle$ follow the adiabatic dynamics of the
Hamiltonian $H_0(t)$. We thus demonstrate that both $H(t)$ and
$\tilde{H}(t)$ can in principle be used to realize STIRSAP
protocol, but $\tilde{H}(t)$ is easier to be manipulated in the
experiments.

\bigskip
\noindent\textbf{Acknowledgements}\\
\noindent  We thank Li You for fruitful discussions. This work was
supported by the NSF of China (Grants No. 11474107, No.
11474153 and No. 11474193), the Shuguang Program (Grant No. 14SG35), the SRFDP
(Grant No. 2013310811003), the Program for Eastern Scholar, the GNSFDYS (Grant No. 2014A030306012), the FOYTHEG
(Grant No. Yq2013050), the PRNPGZ (Grant No. 2014010), the
PCSIRT (Grant No. IRT1243) and SRFYTSCNU (Grant No. 15KJ15).

\bigskip
\noindent\textbf{Author Contributions}\\
\noindent  Y.X.D, Z.T.L., X.X.Y., Q.X.L., W.H., and H.Y. designed
and carried out the experiments. Y.X.D, Z.T.L., Y.C.L., X.C, and
S.-L.Z developed the STIRSAP protocol and performed the numerical
simulations. Y.X.D, X.C., H.Y., and S.-L.Z wrote the paper and all
authors discussed the contents. X.C., H.Y. and S.-L.Z. supervised
the whole project.

\bigskip
\noindent\textbf{Competing Financial Interests}\\
\noindent The authors declare no competing financial interests.

\end{document}